\documentclass[a4paper]{article}
\usepackage{pdfpages}
\pdfoutput=1

\usepackage[utf8]{inputenc}
\usepackage[T1]{fontenc}
\usepackage{upgreek}
\usepackage{nicefrac}
\usepackage{caption}
\usepackage{subcaption}
\usepackage{graphicx}
\usepackage{glossaries}
\usepackage{csquotes}

\usepackage[a4paper, margin=1in]{geometry}

\usepackage{xcolor}
    \usepackage[final]{changes}
\usepackage{ulem}
\usepackage{comment}
\definechangesauthor[name={Stephan}, color={cyan}]{sta}
\definechangesauthor[name={Max}, color={red}]{Max}
    \definechangesauthor[name={Stefan}, color={blue}]{stb}
    \definechangesauthor[name={Stefan}, color={violet}]{stbc}
    \definechangesauthor[name={Stefan}, color={red}]{stbr}
    \definechangesauthor[name={Stefan}, color={orange}]{stbcc}
    \definechangesauthor[name={Stefan Question}, color={orange}]{stbq}

\newacronym{dihm}{DIHM}{digital inline holography}
\newacronym{doah}{DOAH}{digital off-axis holography}
\newacronym{pms}{PMS}{polystyrene micro-sphere}
\newacronym{rbc}{RBC}{red blood cell}
\newacronym{na}{NA}{numerical aperture}
\newacronym{led}{LED}{light emitting diode}
\newacronym{ld}{LD}{laser diode}
\newacronym{usaf}{USAF}{united states air force}
\newacronym{ccd}{CCD}{charge-coupled device}
\newacronym{cmos}{CMOS}{complementary metal-oxide-semiconductor}
\newacronym{tz}{TBY2}{tobacco BY-2 cell}

\date{}
\title{3D-printable portable open-source platform for low-cost lens-less holographic cellular imaging}
\begin{document}
\maketitle
\flushbottom


\begin{center}
\author{\textbf{Stephan Amann$^1$, Max von Witzleben$^1$, Stefan Breuer$^1$}}\vspace{2mm}

$^1$Institute for Applied Physics, Technische Universit\"at Darmstadt, Schlossgartenstra\ss{}e 7, 64289 Darmstadt, Germany
\end{center}
 


\replaced[id=sta]{
Digital holographic microscopy is an emerging potentially low-cost alternative to conventional light microscopy for micro-object imaging on earth, underwater and in space. Immediate access to micron-scale objects however requires a well-balanced system design and sophisticated reconstruction algorithms, that are commercially available, however not accessible cost-efficiently. 
Here, we present an open-source implementation of a lens-less digital inline holographic microscope platform, based on off-the-shelf optical, electronic and mechanical components, costing less than $\$$190. 
It employs a Blu-Ray semiconductor-laser-pickup or a light-emitting-diode, a pinhole, a 3D-printed housing consisting of 3 parts and a single-board portable computer and camera with an open-source implementation of the Fresnel-Kirchhoff routine. 
We demonstrate 1.55$\,\upmu$m spatial resolution by laser-pickup and 3.91\,$\upmu$m by the light-emitting-diode source. 
The housing and mechanical components are 3D printed. Both printer and reconstruction software source codes are open. The light-weight microscope allows to image label-free micro-spheres of 6.5$\,\upmu$m diameter, human red-blood-cells of about 8$\,\upmu$m diameter as well as fast-growing plant Nicotiana-tabacum-BY-2 suspension cells with 50$\,\upmu$m sizes. The imaging capability is validated by imaging-contrast quantification involving a standardized test target. 
The presented 3D-printable portable open-source platform represents a fully-open design, low-cost modular and versatile imaging-solution for use in high- and low-resource areas of the world.}
%
%
{Digital holographic microscopy is an emerging potentially low-cost alternative to conventional light microscopy for micro-object imaging on earth, underwater and in space. Immediate imaging access to micron-scale objects however requires a well-balanced system design and sophisticated reconstruction algorithms, that are commercially available, however not accessible cost-efficiently. 
Here, we present an open-source implementation of a lens-less digital inline holographic microscope based on readily available off-the-shelf optical, electronic and mechanical components, that costs less than $\$$\,190.  
It employs a Blu-Ray semiconductor laser-pickup or alternatively a light emitting diode and an a few micron-diameter pinhole, a 3D printed housing consisting of 3 parts and a single-board portable computer in combination with an open-source implementation of the Fresnel-Kirchhoff routine for micro-scale image reconstruction. 
We experimentally demonstrate spatial resolutions of $1.55\,\upmu$m with a semiconductor laser-pickup and $3.91\,\upmu$m with a light emitting diode as light sources.
The housing and mechanical components are 3D printed and both printer and reconstruction software source codes are open. The light-weight microscope allows to image label-free $6.5\,\upmu$m micro-spheres and human red blood cells with a diameter of about $8\,\upmu$m, as well as fast growing plant Nicotiana tabacum BY-2 suspension cells 
with an individual size of about $50\,\upmu$m. 
We validate the achieved imaging capability by imaging contrast quantification involving a standardized test target. 
The presented 3D-printable portable open-source platform represents a fully open design, low-cost, modular and versatile imaging solution for research laboratories, entry-level imaging in diagnostic labs and student education for use in high- and low-resource areas of the world. 
The open hard and software platform contributes to a democratization of scientific knowledge.}
%
%
\section*{Introduction}
\Gls*{dihm} is an imaging modality within the fast evolving field of imaging microscopy research since a few decades \cite{Jericho_2010} and is based on Gabor's holographic principle \cite{GABOR1948}.
By wide-field illumination of semi-transparent nano- to micron-sized objects with either a coherent or incoherent point light source, an interference pattern forming the hologram is created within a detection plane. It consists of the scattered part of the beam, object beam, and the unscattered (transmitted) part, the reference beam. The hologram contains amplitude and phase information of the imaged objects and allows for a numerical reconstruction of the object's light field. 
%
%
To date, digital holography has been pioneered across a broad range of the electromagnetic spectrum: Terahertz\cite{Rong2015} and infrared~\cite{Repetto2005} holography bear the attractive potential to penetrate opaque media, whereas UV~\cite{Faridian2010,Daloglu2017a} and X-ray~\cite{Krenkel2017,Gorkhover2018} illumination enable imaging on nanometer scale. Electron holography today is a mature research field and widely in use, allowing for molecular imaging \cite{Kreuzer1995,Lichte2007}.
Digital holography within the visible wavelength regime has attracted considerable attention where a broad spectrum of light sources has been employed, ranging from gas \cite{Hobson2013,Molaei2014}, solid state \cite{Carl2004} and semiconductor \cite{Jericho_2010,Giuliano2014,Sanz2017,Rostykus2017} lasers over \glspl*{led} \cite{Mudanyali2010a,Tseng2010,Greenbaum2012,Luo2015} to halogen lamps \cite{Vandewiele2017}. Recently, ultra-broadband digital holography with sunlight illumination has been demonstrated, employing a new reconstruction algorithm \cite{Feng2019}.
%
%
%
In \gls*{doah}, reference and object beam are spatially separated and recombine at an angle within the detector plane. Such experiment can already be compact, combining for example a collimation lens and two gradient-index lenses forming the point-light source and which enables a highly stable off-axis digital holographic system \cite{Serabyn2016}. This helps reducing twin images, typically a challenge in holography set-ups \cite{Leith1962}. During reconstruction a conjugate, out of focus image of the object is obtained in addition to the reconstructed object. This can lead to residual fringes, decreased contrast and an over all reduced image quality. In \gls*{doah}, object and twin image can be separated in the Fourier domain of the hologram. \gls*{doah} set-ups require a rather large
number of optical components including mirrors, beam splitters and collimation lenses.
%
%
Complementary to \gls*{doah}, \gls*{dihm} constitutes a compact holography implementation that avoids to spatially separate the reference beam from the object beam. \Gls*{dihm} setups typically comprise of coherent or incoherent point light source, a 2D digital detector array and, positioned in between, the micron-scale object sample to be imaged. It has been demonstrated in the ultraviolet\cite{Faridian2010}, visible\cite{Carl2004}, near-infrared\cite{Langehanenberg2010} up to the mid-infrared\cite{Ravaro2014} wavelength regime.
%
%
In \gls*{dihm}, only a negligible influence of the twin image on the quality of microscopic object analysis has been demonstrated \cite{Jericho_2010,Kim2010}, as the twin image is defocused across the whole detector array when the distance between detector and object is large as compared to the object size. Moreover, for \gls*{dihm} several approaches to numerically remove the twin image from the reconstructed image have been demonstrated \cite{Latychevskaia2007,Chen2016,Denis2008}.
%
Ideal light sources are semiconductor photonic emitters such as \glspl*{ld} and \glspl*{led} thanks to their compactness, high electro-optical efficiencies, comparable low price and availability. 
%
Partially coherent light sources can even enhance the result in \gls*{dihm} as coherent speckle noise is reduced \cite{Kim2010} while being less susceptible to mechanical vibrations. Considering the generally less complex implementation, easy replacement, lower price, high reliability and reduced safety issues, \glspl*{led} appear as ideal \gls*{dihm} light sources for student and early researcher education.
To ensure homogeneous illumination, the \gls*{led} or \gls*{ld} light can be coupled into an optical fibre ensuring a Gaussian beam profile at the fibre's end \cite{Greenbaum2012,Greenbaum2013,Luo2015,Sanz2017,Frentz2010}. In addition, more compact or complex experimental set-ups can be realized thanks to the mechanical fibre flexibility \cite{Greenbaum2013}.
Both \gls*{ccd} and \gls*{cmos} cameras are commonly used as detectors, with an increasing number of \gls*{cmos} chips due to recent advances in terms of sensitivity and reduced pixel size. 
%
Two parallel tracks of current \gls*{dihm} research can be identified, focusing on two different detection schemes: On-chip microscopy on the one hand, where the sample under investigation is located close to the \gls*{cmos} detector.
Its advantage is the large field-of-view which corresponds to the whole detector area. Several on-chip microscopes employing low-coherent LED illumination \cite{Mudanyali2010a,Greenbaum2013,Scholz2017} have been demonstrated making them already cost-efficient and easy to operate. In combination with several pixel super-resolution approaches using multi-height imaging \cite{Greenbaum2012}, wavelength scanning \cite{Luo2015}, sub-pixel shifting\cite{Bishara2010} or flowing samples \cite{Bishara2010a}, sub-micron spatial resolution is possible. Sub cellular imaging of malaria infected blood cells \cite{Zhu2013}, \gls*{rbc} imaging with a cell phone camera \cite{Tseng2010} and colour imaging using \gls*{dihm} \cite{Greenbaum2013} have been reported. 
On the other hand, fringe magnification technique is studied where the sample is located near the point source \cite{Jericho_2010,Serabyn2016,Sanz2017}. Here, the maximum \textcolor{black}{lateral resolution $\delta_\textnormal{lat}$ is accompanied by a decreased field-of-view (see equation~\eqref{eq:resolution})}.
%

\Gls*{dihm} research has already reached the stage of commercialization where for example several \Gls*{dihm} implementations exist including lens-less inline and \gls*{doah} schemes \cite{lynceetec} delivering lateral resolutions of $0.3\,\upmu$m.
A submersible holographic microscope for remote in-vivo oceanic microscopy has been reported \cite{Jericho2006} and is now commercially available \cite{4deep}.
%
%
Moreover, lens-based \gls*{doah} has recently been employed in in-line industrial production control \cite{holotop} and for imaging of semiconductor structures \cite{Finkeldey2017}.
%
%
In life-sciences, \gls*{dihm} has the advantage of working label-free, which enables non-invasive, in-vivo study of biological samples, for example \glspl*{rbc}
\cite{Park2017,Rappaz2009,Zakrisson2015}, parasited mouse \glspl*{rbc} \cite{Park2015}, sperm cells \cite{Sanz2017}, diarrhea parasites \cite{Mudanyali2010a}, and several aquatic organisms \cite{Jericho2006}. %
Moreover, as information about the whole sample volume is encoded in one single hologram, processing speed is substantially increased as compared to microscopic scanning techniques such as confocal or fluorescence microscopy. Thus, large sample volumes can be analyzed efficiently, such as cell cultures of human cancer cells \cite{Kastl2017,El-Schich2018} as well as microplastic pollution in marine environments \cite{Merola2018}. %
Furthermore, by \replaced[id=stb]{acquiring}{taking} and analyzing sequences of images, \gls*{dihm} enables long term cell evolution studies \cite{Frentz2010} and micro-particle tracking \cite{Choi2009}.
%
%
Although differently complex, portable and low-cost \gls*{dihm} solutions have recently been reported \cite{Dimiduk2010,Tseng2010,Sanz2017,Shimobaba2012}, \replaced[id=stb]{critically important details neccessary for their realization in a laboratory are mostly disclosed. 
Such important details include details on employed}{choice} of light sources, distances between objects and light source or detector, methods of reconstruction, non obvious limitations or design or construction files thus making it challenging to set up a \gls*{dihm} without too much prior expertise and avoids access to low-cost micron spatial resolution imaging. 
%
In this work first, we present two experimental \gls*{dihm} platforms employing an \gls*{led} and a \gls*{ld} as illumination sources and operate the both by a portable single-board computer and camera. The \gls*{ld} has been disassembled from a standard commercially available Blu-ray disk drive. The housing and all mechanical mounts are 3D printed. \cite{raspi:web}. 
Second, we describe the implemented open-source hologram reconstruction based on HoloPy \cite{holopy} and Fiji plugin \cite{Schindelin2012}. All employed code is open-source accessible aiming at triggering further developments and sharing between research laboratories, diagnostic labs and science education.
Third, we demonstrate microscopic 2D-imaging of 
\glspl*{pms} and mature human \glspl*{rbc} imaging with micron spatial resolution. Fourths, we perform 2D-imaging of larger \glspl*{tz}. 
Fifths, we quantify the achieved spatial resolution by optical contrast analysis of an \gls*{usaf} microscopic imaging test target. 
Sixths, we summarize our efforts in developing a 3D-printable open-source platform for cellular imaging and provide a brief outlook on our activities. 
%
%
\subsection*{Laboratory and 3D printed opto-mechanical DIHM setups}
\begin{figure}[b!]
    \centering\includegraphics[width=\textwidth]{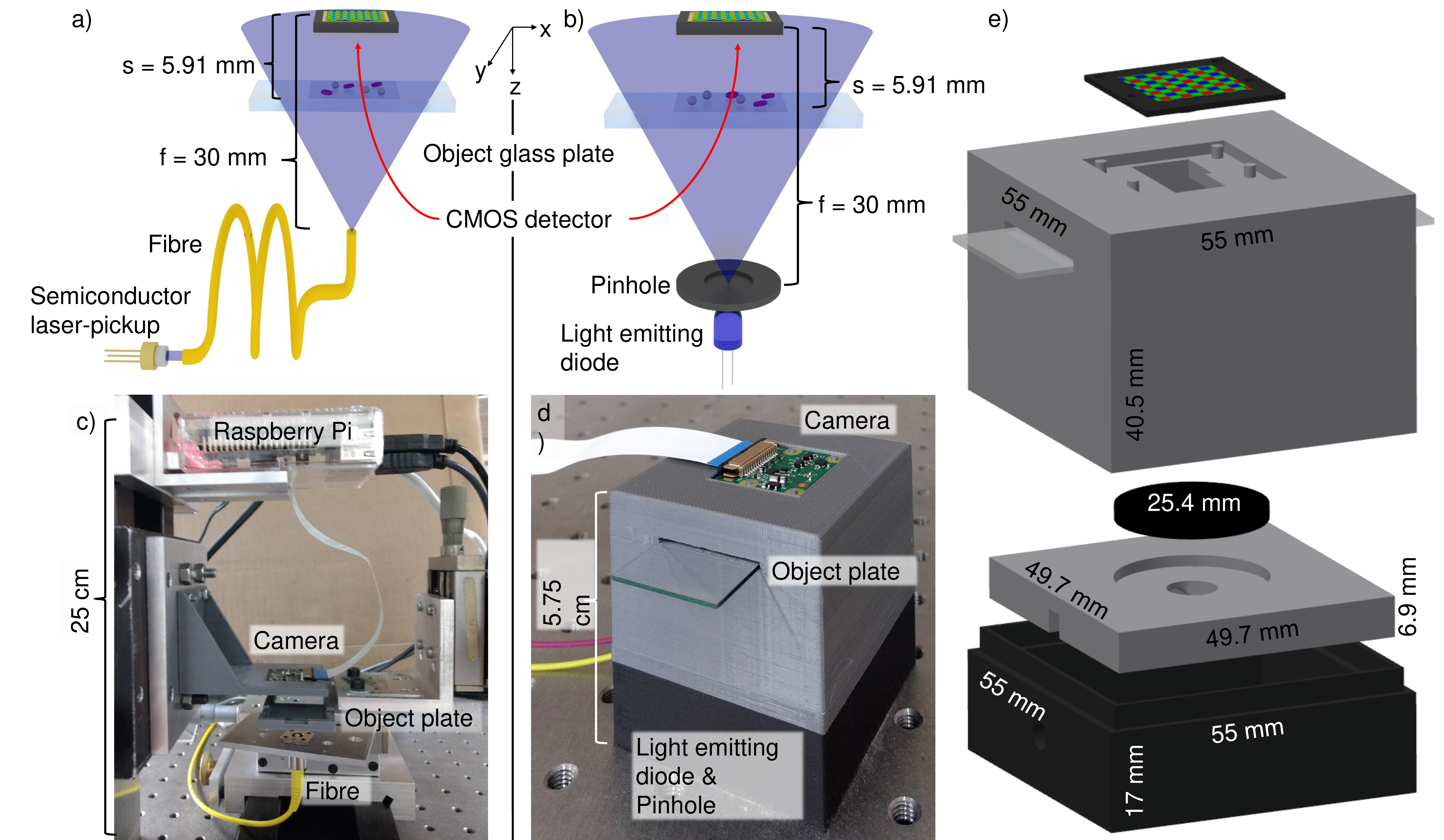}
    \caption{Developed laboratory \gls*{dihm} and 3D printable \gls*{dihm} setup. Schematics of the a) fibre-coupled \gls*{ld} pickup inline illumination. b) \gls*{led} pinhole setup. c) Image of the laboratory fibre-coupled laser-pickup setup and dimensions. d) Image of the complete 3D-printed \gls*{led} set-up. e) Exploded view of the three developed and used 3D printed parts with dimensions (CAD files open accessible, see methods chapter). Schematically depicted light propagation not to scale.}
    \label{fig:set-up}
\end{figure}
%
%
\noindent 
Both developed \gls*{dihm} implementations, described in the following, aim at achieving and validating the maximum achievable spatial resolution when an temporally and spatially coherent source as well as a temporally and spatially incoherent semiconductor light source at equivalent emission wavelength are considered. 
First, a \gls*{ld}-based lens-less \gls*{dihm} platform is developed that aims at \textcolor{black}{demonstrating} 
single-digit micrometer spatial resolution cellular imaging by a temporally coherent source 
and is schematically depicted in Fig.\,\ref{fig:set-up}\,(a). 
%
%
It employs a $405\,$nm emitting \gls*{ld} in a laser-pickup which has been dismounted from a  commercially available standard Blu-ray disc drive. The laser emission is coupled into a standard single-mode fibre. The diverging fundamental Gaussian mode beam is directed towards the object glass plate at a distance of $24.09\,$mm from the fibre exit facet. The \gls*{cmos} camera is positioned at a distance of $f = 30\,$mm.
%
%
Second, an \gls*{led}-based lens-less \gls*{dihm} platform depicted schematically in Fig.\,\ref{fig:set-up}\,(c) is designed and constructed by 3D printable parts, \textcolor{black}{with the over all system costs amounting to less than $\$\,190$.}
Compared to Fig.\,\ref{fig:set-up}\,(a), in Fig.\,\ref{fig:set-up}\,(b) a 430\,nm emitting high-power \gls*{led} is employed as a temporally and spatially incoherent semiconductor light source. A fraction of the emitted light is passed through a high-precision pinhole 
where 1.1\,$\upmu$W of optical power are emitted at the pinhole for an injection current of 125\,mA and impinge on objects to be imaged ($z=$5.91\,mm).
\textcolor{black}{For live cell imaging, such ultra-low optical power is of critical importance, as cell damage by light exposure needs to be minimised.}
To construct the \gls*{led}-based platform in Fig.\,\ref{fig:set-up}\,b), first three mechanical parts in Fig. \ref{fig:set-up}\,c) are 3D-printed, see section \enquote{Methods},
and assembled as sketched in Fig.\,\ref{fig:set-up}\,d) and e) where also specific dimensions of the platform are depicted.
%
%
For both experiments, equal spatial separations between emission facet, microscope glass plate carrying the micro-objects under investigation and \gls*{cmos} detector are chosen. \textcolor{black}{\added[id=sta]{We chose a distance of 30\,mm between light source and detector in order to reach a compact experimental set-up, while still maintaining illumination of the hole detector area. The position of the object emerges from resolution optimization, see section \enquote{Resolution}.}}
%
%
A Raspberry Pi single-board portable computer 
and a Raspberry Pi \gls*{cmos} detector camera 
with a pixel size of $(1.12 \times 1.12)\,\upmu$m$^2$ serve as light source current injection driver and hologram acquisition, see section \enquote{Methods}.
A constant injection current of $28\,$mA for the \gls*{ld} were provided by a commercial \gls*{ld} driver, while a current of $125\,$mA for the \gls*{led} was provided by a simple electrical circuit made of off the shelf components.
%
%
A conventional rechargeable power bank battery pack provides electrical energy for the computer.
Estimated theoretical spatial resolutions $\delta$ of $\delta_{\textnormal{Laser}}=0.87 \, \upmu$m and with an \gls*{led} light source of $\delta_{\textnormal{LED}}=0.92 \, \upmu$m at an available \gls*{cmos} pixel size of 1.12\,$\upmu$m and at a distance of $f=$30\,mm between fibre facet and \gls*{cmos} detector are expected for the \gls*{ld} and \gls*{led} light source, respectively.
%
\subsection*{Image reconstruction}
Following the hologram acquisition, information retrieval of the cellular objects deposited on the object glass plate is performed numerically. The propagation of light fields is completely described by diffraction theory. Hence it is possible to reconstruct amplitude and phase information of the objects from their interference patterns generated on the camera.
At the object location the pattern is focused and reveals the shape and morphology of micro-objects. 
Numerically, arbitrary planes can be re-focused in retrospect yielding access to volumes with a large number of objects in different heights in z-direction which can be studied with acquiring a single image. 
The propagation of a wave front towards the detector is described by the Fresnel-Kirchhoff diffraction integral
\begin{equation}
    U_\textnormal{det}(X,Y)=-\frac{i}{\lambda}\int\int U_\textnormal{in}(x,y)t(x,y) \frac{\exp{\left(ik\lvert \vec{r}-\vec{R} \rvert\right)}}{\lvert \vec{r}-\vec{R}\rvert }dx dy,
\end{equation} 
\noindent  where $U_\textnormal{det}(X,Y)$ denotes the wave field on the detector, $U_\textnormal{in}(x,y)$ is the incident wave, $t(x,y)$ the transmission function of the object, and $\vec{r}=(x,y,z)$ and $\vec{R}=(X,Y,0)$ are two points in the object plane respectively detector plane\cite{Latychevskaia2015}.
The amplitude and phase distribution of the object can be obtained via an inverse integral
\begin{equation}
    U_\textnormal{obj}(x,y)=\frac{i}{\lambda}\int\int U_\textnormal{ref}(X,Y)H_\textnormal{det}(X,Y) \frac{\exp{\left(-ik\lvert \vec{r}-\vec{R} \rvert\right)}}{\lvert \vec{r}-\vec{R}\rvert }dX dY,
\end{equation} 
\noindent with the reference wave $U_\textnormal{ref}(X,Y)$ and the intensity distribution of the hologram in the detector plane $H_\textnormal{det}(X,Y)$. For the reconstruction, an open-source plugin for the open-source microscope software Fiji, see \enquote{Methods}~section,
that implements the angular spectrum estimation for small distances in the order of micrometers up to several centimeters, is employed. The amplitude distribution in the reconstructed plane is calculated using two Fourier transforms
\begin{equation}
    U_\textnormal{obj}(z)=\mathcal{F}^{-1} \Bigg \{  \mathcal{F}\{U_\textnormal{det}\}\exp{\left( iz\sqrt{k^2-\frac{4 \pi^2n^2}{N p}} \right) } \Bigg \},
\end{equation} 
where $z$ is the height of the reconstructed plane, $k$ the wave vector, $n$ the index of refraction, $N$ the number of pixels and $p$ the pixel size of the CMOS detector. $\mathcal{F}$ and $\mathcal{F}^{-1}$ denote Fourier Transform and the inverse Fourier Transform. 
%
The formalism described above is implemented in two open-source reconstruction software packages, see \enquote{Methods}~section.   
In the following, we elaborate and identify two easy to implement reconstruction softwares on a standard desktop computer or potentially also on a mobile phone.
%
HoloPy, a software package for python, allows for hologram reconstruction, but also hologram simulation and scattering calculations.
The algorithm to reconstruct point source holograms is based on Fresnel-Kirchhoff diffraction\cite{Jericho_2010}. It considers a background subtracted hologram, experimental parameters including distances and light source wavelengths and then reconstructs the hologram using two Fourier transforms. Alternatively, hologram fitting is provided by HoloPy where the position of a scatterer is simulated in order to produce the same interference pattern instead of image back-propagation \cite{Dimiduk2014}. For the case of a known number of scatterers, this method can be recommended as it allows to reconstruct spherical or cylindrical object shapes. It is less practicable, however, when arbitrarily shaped micro-objects are of interest, as for example folded \glspl*{rbc}.
%
For the latter case, an open-source plugin\,\cite{Garcia-Sucerquia2015} for 
Fiji, see \enquote{Methods}~section, is a possible solution with a user-friendly graphical-user-interface, implemented in Java. Phase, amplitude and intensity distribution of micro-objects can be reconstructed at arbitrary heights in z-direction.
\subsection*{Micro particle and cellular imaging}
\begin{figure}[ht!]
\centering\includegraphics[width=\textwidth]{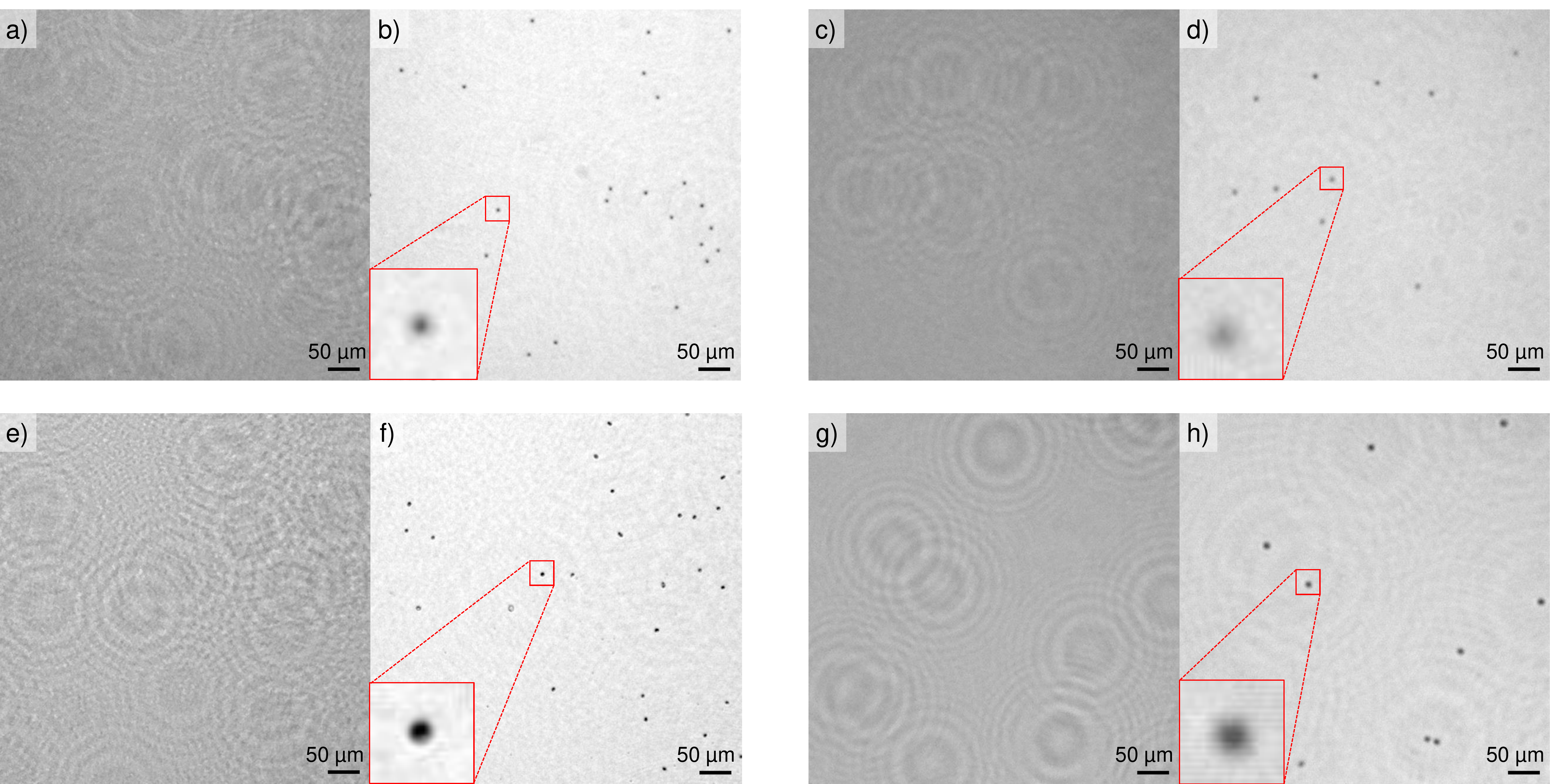}
\caption{Holograms and according reconstructions of \glspl*{pms} and human \glspl*{rbc}. Each inset shows a single sphere or \gls*{rbc} enlarged to five times its original size. \textcolor{black}{a) Hologram of \glspl*{pms} captured with \gls*{ld} setup. b) Reconstruction of (a). c) Hologram of \glspl*{pms} captured with \gls*{led} setup. d) Reconstruction of (c). e) Hologram of \glspl*{rbc} captured with laser setup. f) Reconstruction of (e). g) Hologram of \glspl*{rbc} captured with \gls*{led} setup. h) Reconstruction of (g).}}
\label{fig:Micro_objects}
\end{figure}
In the following, we first capture and image standardized \glspl*{pms} of diameter $(6.5\pm 0.2)\,\upmu$m by both the \gls*{ld} and  \gls*{led}-based lens-less \gls*{dihm} platform and reconstruct the resulting object properties by the Fiji plugin. Second, we investigate anonymized mature \glspl*{rbc} as micro-objects in the same manner. Third, we image cell suspension culture \glspl*{tz}. 
%
%
The recorded holograms and subsequently reconstructed object planes for multiple \glspl*{pms} and \glspl*{rbc} are depicted in Fig.\,\ref{fig:Micro_objects}. Laser-based \gls*{dihm} hologram (a) and reconstruction (b) is presented next to \gls*{led}-based \gls*{dihm} hologram (c) and reconstruction results (d). 
Both insets depict an isolated \gls*{pms} or \gls*{rbc} enlarged to five times its original size.
%
It becomes evident that the \gls*{ld}-based platform provides sharper images where a larger number of interference fringes can be captured per object.
These fringes overlap within the hologram resulting in a hologram with more grainy texture as compared to the \gls*{led}-based results in Fig.\,\ref{fig:Micro_objects}\,(a,c). In contrast, the \gls*{led}-based reconstructed image is considerably more washed out resulting in comparably extended objects.
Accordingly, for human \glspl*{rbc} imaged by the \gls*{ld}-based platform, the oval 
disk shape can clearly be resolved as depicted in Fig.\,\ref{fig:Micro_objects}\,(e,f). Several \glspl*{rbc} appear to be tilted in their spatial position, resulting in an elliptical shape. 
This is in stark contrast to the results obtained by the \gls*{led}-based platform where information retrieval, for example on the cell morphology, are scarce. However, by the \gls*{led}-based platform, individual cells can clearly been distinguished, thus exemplifying its potential for individual cell counting or tracking.
\begin{figure}[ht!]
\centering\includegraphics[width=\textwidth]{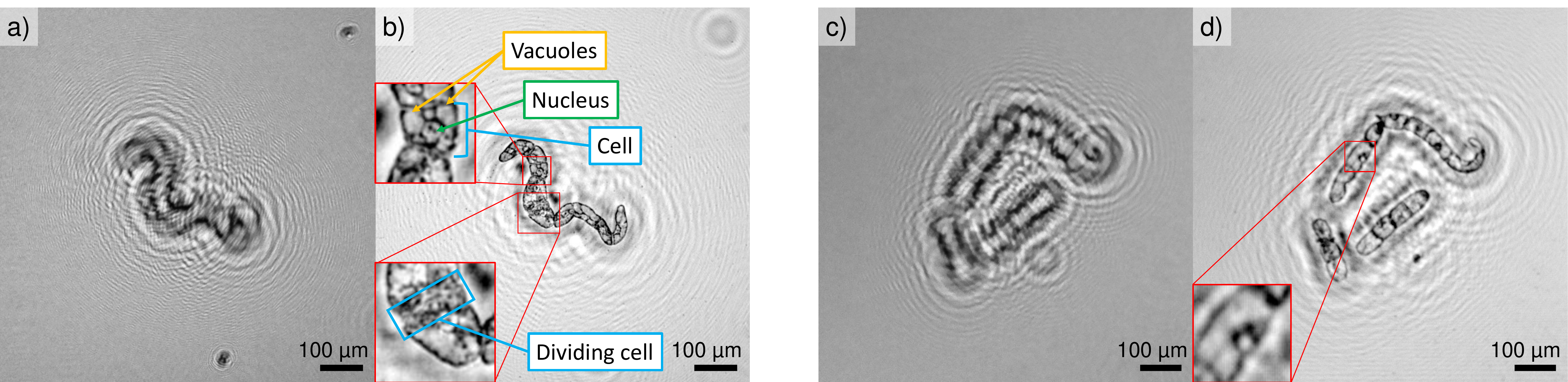}
\caption{Holograms and according reconstructions of \glspl*{tz}. \textcolor{black}{The insets are depicted magnified by a factor of three.} a) Hologram 
captured with \gls*{ld} setup. b) Reconstruction of (a). c) Hologram 
captured with \gls*{led} setup. d) Reconstruction of (c).}
\label{fig:Tabak}
\end{figure}
%
%
%
In order to validate both platform's imaging capabilities also for extended cellular objects, fast growing plant tobacco \glspl*{tz} have been prepared and imaged. \glspl*{tz} are employed in various fields of plant biology as a model material and are ideally suited for cellular and molecular analyses\cite{Yokoyama2004}. 
Corresponding results are depicted in Fig.\,\ref{fig:Tabak}. 
The recorded holograms and reconstructed object planes for an isolated \gls*{tz} are depicted for \gls*{ld}-based DIHM hologram (a) and reconstruction (b) is presented whereas the hologram, obtained by the \gls*{led}-based \gls*{dihm}, is depicted in (c) and the corresponding reconstruction in (d).
Both platforms allow to successfully access individual cell segments with a length of $50\,\upmu$m as well as internal structures including cell nuclei and vacuoles. In Fig. \ref{fig:Tabak}b), a dividing cell undergoing mitosis can be observed. For \gls*{led} illumination, individual vacuoles are not distinguishable. This is not surprising, as the vacuole membrane thickness is around one order of magnitude smaller than the cell wall thickness of (7-10)\,nm for plant vacuoles\cite{Esau1962} as compared to (71-87)\,nm for tobacco leaf cells walls \cite{Hoffmann-Benning1997}. However it is possible to make out the nuclei of several cells. Interestingly, \glspl*{tz} infer a more complex interference pattern as compared to both \glspl*{pms} and \glspl*{rbc}, indicating a stronger absorption and thus increased hologram contrast. 
We found that Fiji revealed a substantially faster reconstruction as compared to HoloPy. The reconstruction of 10 planes of a digital hologram by the Fiji plugin demands 30 seconds computational time on a regular consumer PC as compared to several minutes by HoloPy. Towards larger volumes, the reconstruction time can theoretically be improved by performing computations on a graphics processing unit\,\cite{Hobson2013} as demonstrated for live imaging\,\cite{Locatelli2015}.
In the following section, we aim to quantify the theoretical lateral resolution as well as the spatial resolution experimentally achieved by both the \gls*{ld}-based and \gls*{led}-based \gls*{dihm} platforms.
\subsection*{Resolution}

In \gls*{dihm}, the lateral resolution is bounded by the optical assembly numerical aperture ($N\!A$) and the illumination wavelength $\lambda$ 
\cite{Jericho_2010}: 
\begin{equation}
    \delta_\textnormal{lat} = \frac{\lambda}{2N\!A},
\end{equation}
suggesting a shorter wavelength for a higher lateral resolution. For a digital holographic microscope with a pixel number $N$, with pixel size $p$, an illumination wavelength $\lambda$ and a distance $s$ between object and detector plane, this translates to 
\begin{equation}
\label{eq:resolution}
    \delta_\textnormal{lat} = \frac{s \lambda}{N p}.
\end{equation}
As described in \cite{Serabyn2017}, a maximum lateral resolution can be achieved at a distance 
\begin{equation} \label{eq:best distance}
    s_\textnormal{opt}=\frac{f}{1+\frac{f \lambda}{N p^2}}
\end{equation} 
between object and detector, with the distance $f$ between light source and detector.
This originates from the consideration, that a higher resolution is possible, the closer the object is placed to the detector, however at the same time the interference fringes move closer. Here, $s_\textnormal{opt}$ denotes the distance at which different interference rings are still resolved by different camera pixels. The axial resolution of the system can be calculated according to
\begin{equation}
    \delta_\textnormal{ax} = \frac{\lambda}{2(N\!A)^2} =  \frac{2s^2 \lambda}{(N p)^2}.
\end{equation}
\begin{figure}[!ht]
\centering\includegraphics[width=\textwidth]{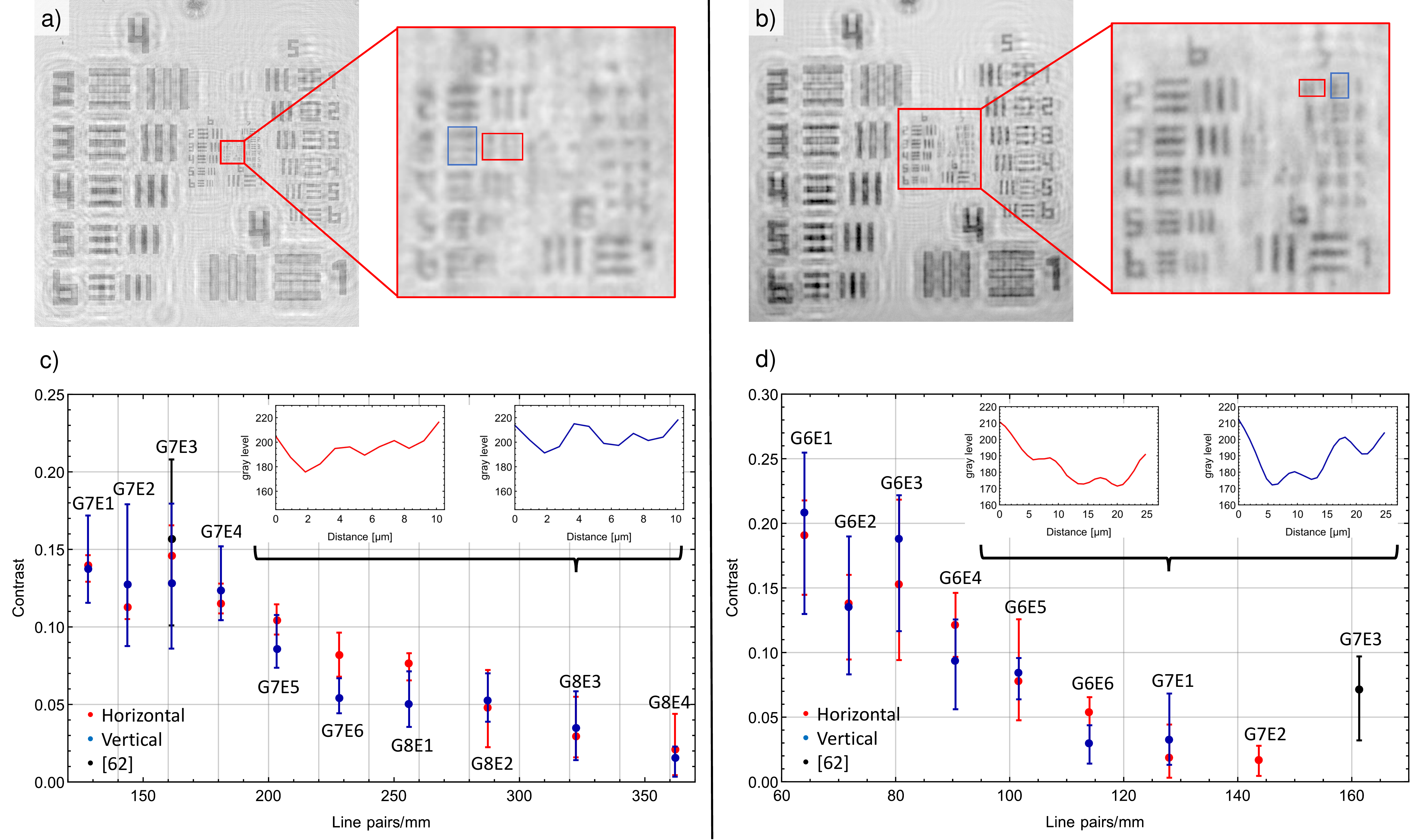}
\caption{Imaging of a 1951 \gls*{usaf} target to quantify the accessible spatial resolution. a) \gls*{ld} setup. Groups 8 and 9 are shown enlarged. b) \gls*{led} setup. Groups 6 to 9 are shown enlarged. c) Contrast of the different elements for \gls*{ld} illumination. The inset shows the intensity profile plot for horizontal and vertical elements 3 of group 8. d) Contrast for \gls*{led} illumination. 
Additionally included are reported contrast values \cite{Feng2017}, where either an unspecified laser or an \gls*{led} emitting at 470\,nm ($25\,\upmu$m pinhole) had been employed (not specified further).}
\label{fig:USAF}
\end{figure}
\noindent By equation~\eqref{eq:best distance}, the theoretical resolution for both developed platforms can be estimated. For $p=1.12\,\upmu$m, $N=2464$, $f=30\,$mm, $\lambda_\textnormal{LED} = 430\,$nm  and  $\lambda_\textnormal{Laser} = 405\,$nm, respectively, lateral resolutions of $\delta_\textnormal{LED}=0.92\,\upmu$m and $\delta_\textnormal{Laser}=0.87\,\upmu$m are theoretically possible by the selected, optimum platform design.
To evaluate the experimentally achieved resolution, a 1951 \gls*{usaf} microscopic imaging test target on a glass microscopic slide serves as a reference object
consisting of groups of horizontal and vertical lines with decreasing spatial frequency. 
The resulting reconstructed amplitude images acquired with the \gls*{led} (a) and \gls*{ld} (b) setup are depicted in Fig.\,\ref{fig:USAF}.
With \gls*{led} illumination, element 1 of group 7 is the last resolvable element corresponding to a resolution of 128 line pairs/mm and a line width of 3.91\,$\upmu$m.
For \gls*{ld} illumination, element 3 of group 8 is still resolvable, leading to a resolution of 322.5 line pairs/mm and a line width of 1.55\,$\upmu$m. This is mostly a result of the higher temporal and spatial resolution of the laser in comparison to the used LED, as well as the smaller wavelength. The resolving power with \gls*{ld} illumination is thus significantly higher.

Finally, to quantify the achieved image contrast and thus evaluate the capability of the developed \gls*{dihm} setups, in the following the intensity of a single \gls*{usaf} element is averaged along its axis and plotted (see red and blue rectangle in Fig. \ref{fig:USAF}a and b). Then, the contrast of consecutive extreme points is calculated by $K=\frac{I_\textnormal{max}-I_\textnormal{min}}{I_\textnormal{max}+I_\textnormal{min}}$. Each element consists of 5 dark lines which leads to 4 contrast values. The average contrast is then displayed in dependence on the resolution in line pairs/mm, with the error bars ranging from the minimum to the maximum value. 
The resulting image contrast of different elements of the respective \gls*{usaf} image is depicted in Fig.\,\ref{fig:USAF}c) and d).
Results obtained by both light sources indicate decreasing contrast values with increasing line pairs. The maximum resolvable elements are as shown by the the intensity profile plots for horizontal and vertical elements 3 of group 8 in the inset of Fig.\,\ref{fig:USAF}c). For \gls*{led}, the maximum resolvable element is element 1 of group 7 as depicted in the inset of Fig.\,\ref{fig:USAF}d). We compared our contrast values to reported values \cite{Feng2017}, where an unspecified laser or a \gls*{led} with a central wavelength of 470\,nm (25\,$\upmu$m pinhole) was used for image acquisition. The reported value for laser illumination coincides with our results. 
With \gls*{led} illumination, the reported value reachs a higher resolution, group 7 element 3 is still resolvable. This could be due to a higher signal-to-noise ratio of the employed camera, the smaller bandwidth of only 10\,nm as well as the use of advanced reconstruction algorithms.
%
The achieved spatial resolution of both setups is suited to image and detect individual micro-particles and cellular objects including ensembles of microscopic biological samples. For the \gls*{ld}-based \gls*{dihm}, sharper interference pattern and a higher number of interference fringes could be captured. Thus, more object information is retrieved yielding a crisper reconstructed image including more details. We attribute this to the \gls*{ld}'s higher second-order temporal coherence as well as spatial coherence. The \gls*{led}'s spatial coherence could be increased by reducing the pinhole diameter which would require increased \gls*{led} biasing currents which then require efficient cooling of the \gls*{led} and thereby increasing the physical dimensions of the setup. 
Our \glspl*{dihm} allow to image cellular objects with dimensions smaller than $10\,\upmu$m as well as microscopy of larger objects with a spatial resolution of $3.91\,\upmu$m. To ensure maximum possible resolution, a \gls*{ld}-based \glspl*{dihm} is recommended, whereas for applications where high temporal stability, compact set-up, ruggedness and reduced costs a required, \gls*{led}-based \glspl*{dihm} is recommended.
%
\section*{Conclusion}
A 3D printable platform for lens-less holographic cellular imaging with open accessible software solutions has been developed delivering spatial resolutions of $1.55\,\upmu$m by \gls*{ld} or $3.91\,\upmu$m by \gls*{led} illumination. 
A 405\,nm Blu-ray semiconductor laser-pickup coupled to an optical fibre and a 430\,nm high power \gls*{led} in combination with a $15\,\upmu$m pinhole have been successfully employed as \gls*{dihm} light sources. 
Despite its lower degree of temporal coherence, the \gls*{led} proved to be of advance in terms of implementation, price and lower safety concerns.
A single-board portable Raspberry Pi computer and camera 
operate the light sources as well as perform the image acquisition. 
By an open-source software implementation of the Fresnel-Kirchhoff algorithm, we imaged and succesfully reconstructed $6.5\,\upmu$m \gls*{pms} and human \glspl*{rbc} with a diameter of about $8\,\upmu$m, as well as \glspl*{tz} with an individual size of about $50\,\upmu$m.
Less than \textcolor{black}{$1.1\,\upmu$W} of optical power were sufficient for holographic imaging microscopy. Such ultra-low optical power can be of critical importance for live cell imaging where light exposure of the cells needs to be as low as possible.
Equally compact setup could be envisioned for the \gls*{ld} when fibre-coupled \glspl*{ld} are available, which are however considerably expensive.
The \gls*{dihm} setup presented here may serve as a reliable, easy to implement and flexible to extend solution for student an early researcher education and for different demands in microscopic imaging.
The total costs for the \gls*{led} setup 
amount to \$\,190 (\$\,3 \gls*{led}, \$\,75 Pinhole , \$\,35 Raspberry Pi~3, \$\, 25 Raspberry Pi Cam~v2, \$\,25 3D print, \$\,27 power bank) and thus enables a convenient entry into the wide field of digital holography. 
Future work could include the integration of the DIHM with micro-fluidic channels \cite{Choi2009,Cheong2017,Bianco2017} or considering machine learning algorithms to automatically count and identify particles \cite{Roy2014}, as well as diagnose illnesses such as meningitis \cite{Delacroix2017}, iron-deficiency anemia or diabetes mellitus \cite{Kim2019}.
The developed 3D-printable photonic platform might help facilitating reproducibility of results obtained in different laboratories and prototyping of specific improvements and advancements of the \gls*{dihm} setups. The cost-efficient open science and open hard and software platform aims in particular at contributing towards a democratization of scientific knowledge \cite{Baden2015}.
\section*{Methods}
\subsection*{DIHM construction, light sources, opto-electronics, electrical circuits and 3D print}\label{methods:construction}

\begin{figure}[ht!]
    \centering\includegraphics[width=\textwidth]{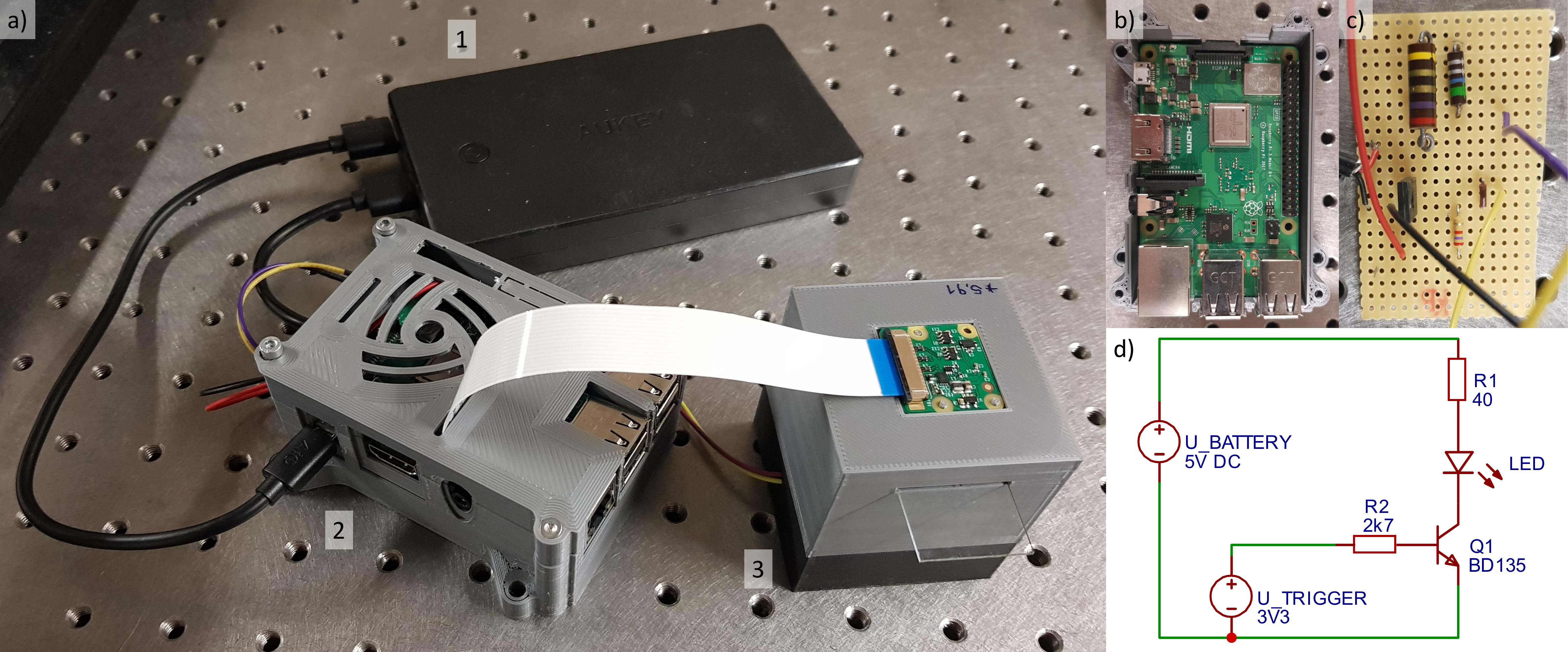}
    \caption{a) The fully portable DIHM system consisting of 1. USB-port power bank, 2. 3D printed housing containing Raspberry Pi and LED control circuit, 3. LED DIHM system as presented in Fig. \ref{fig:set-up}d). \textcolor{black}{b) Closeup of the Raspberry Pi single-board computer.} c) Closeup of the LED control circuit. d) Schematic of the circuit for biasing the LED using the Raspberry Pi's GPIOs.}
    \label{fig:Circuit+Batt}
\end{figure}

The Blu-ray \gls*{ld}-pickup 
(SF-AW210) has been dismounted from a commercially available standard computer Blu-ray optical head and its emits a maximum optical output power of $300\,$mW at a wavelength of 405\,nm for an injection current of $150\,$mW.
The \gls*{ld}-pickup beam is collimated and focussed by two aspheric lenses with an effective focal length of $11\,$mm ($\$\,87$ each, C220TMD-A, Thorlabs Inc.) 
into a $2$\,m long silica core FC/PC single-mode optical fibre with a core diameter of $3\,\upmu$m ($\$\,99$, S405-XP, Nufern Inc.)
%
the coupling is adjust to ensure a low laser power of $24.5\,\upmu$W at the fibre exit which proved optimum for DIHM imaging. The emitted divergent beam (Rayleigh length $z_\textnormal{r}= 21\,\upmu$m) has TEM$_{00}$ intensity profile and a mode field diameter of $2 w_0 = 3.3\,\,\upmu$m.
Within the plane of the object glass plate, at z = 5.91\,mm, the divergent beam impinges the objects under investigation with a power density of 4.2\,$\upmu$Watt/mm$^2$ and with a beam diameter of 3.8\,mm (1/e$^2$). In the CMOS detector plane ($f = $30\,mm), the beam diameter amounts to 4.7\,mm (1/e$^2$).
The high-power \gls*{led} ($\$\,3$, 3W High Power LED 430nm - 435nm hyper violet, Avonec, Germany \cite{avonec:web})
emits at wavelengths centered at around $430$\,nm with spectral bandwidth of $\Delta \lambda_\textnormal{LED}$ of 15\,nm (full-width-half-maximum) corresponding to a coherence length of $L_\textnormal{c,LED} = \lambda_\textnormal{LED}^2 \times (\pi\times\Delta \lambda_\textnormal{LED})^{-1} = 
3.9\,\upmu$m for a Gaussian intensity distribution, with $\Delta \lambda$  the full width at half-maximum spectral line width\cite{Mandel1995}.
%
A high-precision stainless steel pinhole with a diameter of (15\,$\pm 1.5)\,\upmu$m ($\$\,75$, P15D, Thorlabs Inc.) has been employed as all 3D printing attempts yet did not yield a necessary sufficiently high mechanical grade circular diameter.
All neccessary parts for the \gls*{led} setup can be found at \url{https://github.com/teph12/DIHM}. All parts and Raspberry Pi housing were printed with a commercial 3D-printer ($\$\,880$, Prusa i3 MK3, Prusa Research s.r.o.) using standard polylactide synthetic polymer filament with 1.75\,mm diameter. A spatial printing resolution of 0.4\,mm, parallel to the optical table, and 0.05\,mm, in vertical or z-direction is available. Within the printed \gls*{dihm} housing, \gls*{led} and pinhole were fixed with tape ($\$\,9$, 3M-ID 70005241826 Scotch Magic Tape, 3M Inc.). The cased with VESA mounts for Raspberry Pi 3 (B/B+), Pi 2 B, and Pi 1 B+ can be accessed by\cite{case:web}.
The \gls*{cmos} camera module is glued to the upper part of the box 
after removing the  lens mounted in front of the module. We observed that otherwise strong hologram distortions appeared.
%
In order to electrically bias the \gls*{ld}, a commercial \gls*{ld} driver was used to provide a constant output current. However, an open-source driver is under construction while all parts are available for in total $\$\,20$\cite{web:Innolasers}. For the \gls*{led}, a custom soldered circuit has been developed accessing Raspberry Pi's general purpose input/output (GPIO). The assembled system is depicted in \ref{fig:Circuit+Batt}b). The driver circuits can well be integrated into the 3D printed Raspberry Pi housing. 
A diagram of the circuit can be seen in Fig. \ref{fig:Circuit+Batt}a). 
The portable computer, camera and \gls*{dihm} light sources are supplied with electrical power by a conventional power bank battery pack with maximum output power of 10\,W \textcolor{black}{(\$27, Aukey PB-N36)}. A computer monitor and computer mouse are required for the hologram acquisition.
V\_BATTERY is the voltage provided by the power bank, whereas V\_TRIGGER is the voltage between the Raspberry Pi GPIO and ground. It triggers a transistor (BD135), which lets a current of 125\,mA flow through the LED. 
\subsection*{Light source biasing, image detection and aquisition}\label{methods:biasing}
A portable single-board computer and camera module with a 3280 x 2464 pixel CMOS chip and pixel dimensions of ($1.12\times1.12)\,\upmu$m$^2$, sensor size ($3.674 \times 2.760$)\,mm$^2$  ($\$\,35$, Raspberry Pi 3 with $\$\,25$ Raspberry Pi Camera v2, Raspberry Pi Foundation, UK \cite{raspi:web}) serve as light source driver and hologram acquisition.
\subsection*{Sample preparation}\label{methods:preparation}
The RBC samples under investigation are standard anonymized \gls*{rbc} reference cells ($\$\,27$, A1A2B0 4x10, Immucor Med. Diagnostik GmbH, Germany). They are diluted in saline solution to a high degree (up to 0.02\,\%). The employed polystyrene micro-spheres ($\$\,129$, BS-Partikel GmbH, Germany) have a diameter of $(6.5 \pm 0.2)\,\upmu$m. They are equally high diluted in distilled water. Furthermore a few ml of ordinary dish detergent are added to prevent agglomeration and adhesion of the micro-spheres. After diluting the particular object a drop of the solution is placed by a standard plastic pasteur pipette on a high transmission flat glass microscope slide (75\,mm x 25\,mm x 1\,mm, B270 I, SCHOTT AG) and then covered with a glass cover plate (22\,mm x 22\,mm x 0.15\,mm, B270 I, SCHOTT AG)). 
Nicotiana tabacum cv. BY-2 suspension plant cell cultures\cite{Nagata1992} were grown in liquid saline medium based on a modified Linsmaier and Skoog medium with agitation on a incubator shaker. 
The cells have been grown in 50\,ml medium within a 250\,ml glass flask. Every 7 days, 5\,$\%$ inoculum had been transferred into a fresh medium
\cite{Nocarova2009,Mercx2016} and stored permanently on an incubator shaker. 
\subsection*{Image acquisition, reconstruction and resolution validation}\label{methods:acquisition}
The pictures are acquired with a fixed white balance. For details see the camera file on \url{https://github.com/teph12/DIHM}. The image is then transferred to a PC with Windows 10 operating system and equipped with an Intel i3 processor and 8\,GB of RAM, 
with Fiji installed. In Fiji the image is converted to a 32-bit black and white picture, which is then loaded in the \enquote{Numerical Reconstruction} plugin. Here, using the parameters of image acquisition (distance between camera and object, wavelength, image size) the image is reconstructed. Subsequently the image contrast is normalized and enhanced by 0.2\,\%. 
For the quantification of the experimentally achieved spatial resolution, a commercially available standardized 1951 \gls*{usaf} positive high-contrast chrome on quartz glass microscopic imaging test target created by photo lithography on a glass microscopic slide serves as a reference object ($\$$\,900, Ready Optics, US). It consists of groups of horizontal and vertical lines with standardized spatial frequencies starting at group 4, element 1 with 31\,$\upmu$m spacing and ending at group 11, element 6 with 137\,nm spacing.

\bibliography{Literaturliste}
\bibliographystyle{unsrt}

\end{document}